\documentclass[aps,prd,twocolumn,superscriptaddress,nofootinbib,10pt]{revtex4-2}

\usepackage{amsmath,amssymb,mathrsfs,bm}
\usepackage{amsthm}
\usepackage{graphicx}
\usepackage[colorlinks=true,linkcolor=blue,citecolor=blue,urlcolor=blue]{hyperref}
\usepackage{tikz}
\usepackage{xcolor}

\newtheorem{proposition}{Proposition}
\newtheorem{theorem}{Theorem}

\newcommand{\Kerr}{\mathrm{K}}
\newcommand{\Sexc}{S_{\mathrm{exc}}}
\newcommand{\Sgeom}{S_{\mathrm{geom}}^{(2)}}
\newcommand{\Sgeomreg}{S_{\mathrm{geom,reg}}^{(2)}}
\newcommand{\SgeomKerr}{S_{\mathrm{geom,K}}^{(2)}}
\newcommand{\Sbook}{S_{\mathrm{book}}}
\newcommand{\etaexc}{\eta_{\mathrm{exc}}}
\newcommand{\mtilde}{\widetilde m}

\newcommand{\Sexcdot}{\dot S_{\mathrm{exc}}}

\definecolor{lime}{HTML}{A6CE39}
\DeclareRobustCommand{\orcidicon}{%
	\begin{tikzpicture}
		\draw[lime, fill=lime] (0,0) 
		circle [radius=0.16] 
		node[white] {{\fontfamily{qag}\selectfont \tiny ID}};
		\draw[white, fill=white] (-0.0625,0.095) 
		circle [radius=0.007];
	\end{tikzpicture}
	\hspace{-2mm}
}

\foreach \x in {A, ..., Z}{%
	\expandafter\xdef\csname orcid\x\endcsname{\noexpand\href{https://orcid.org/\csname orcidauthor\x\endcsname}{\noexpand\orcidicon}}
}

\newcommand\orcidFrancisco{{\href{https://orcid.org/0000-0002-9388-8373}{\orcidicon}}}
\newcommand\orcidJorge{{\href{https://orcid.org/0000-0002-4988-7891}{\orcidicon}}}
\newcommand\orcidManuel{{\href{https://orcid.org/0000-0001-8586-0285}{\orcidicon}}}

\begin{document}

\title{Excess inner-horizon entropy and the thermodynamics of Kerr formation}

\author{Francisco S. N. Lobo\orcidFrancisco\!\!} 
\email{fslobo@ciencias.ulisboa.pt}
\affiliation{Instituto de Astrof\'{i}sica e Ci\^{e}ncias do Espa\c{c}o, Faculdade de Ci\^{e}ncias da Universidade de Lisboa, Edifício C8, Campo Grande, P-1749-016 Lisbon, Portugal}
\affiliation{Departamento de F\'{i}sica, Faculdade de Ci\^{e}ncias da Universidade de Lisboa, Edif\'{i}cio C8, Campo Grande, P-1749-016 Lisbon, Portugal}
\author{Jorge Ovalle\orcidJorge\!\!} 
\email{jorge.ovalle@physics.slu.cz}
\affiliation{Research Centre for Theoretical Physics and Astrophysics, Institute of Physics, Silesian University in Opava, CZ-746 01 Opava, Czech Republic}
\affiliation{Universidad de Tarapac\'{a}, Avenida Luis Emilio Recabarren 2477, Iquique, Chile}
\author{Manuel E. Rodrigues\orcidManuel\!\!} 
\email{esialg@gmail.com}
\affiliation{Faculdade de F\'{i}sica, Programa de P\'{o}s-Gradua\c{c}\~{a}o em F\'{i}sica, Universidade Federal do Par\'{a}, 66075-110, Bel\'{e}m, Par\'{a}, Brazil}
\affiliation{Faculdade de Ci\^{e}ncias Exatas e Tecnologia, Universidade Federal do Par\'{a}, Campus Universit\'{a}rio de Abaetetuba, 68440-000, Abaetetuba, Par\'{a}, Brazil}

\date{\today}

\begin{abstract}
We develop a charge-matched thermodynamic bookkeeping framework for the regular rotating black-hole family underlying an exact model of Kerr formation.  Since the Kerr endpoint of the interpolating trajectory already contains an inner Cauchy horizon, the rotational analogue of the hidden entropy of regular Schwarzschild black holes is not the full inner-horizon area term, but the excess geometric contribution relative to the Kerr inner horizon with the same mass and angular momentum.  For a stationary regular configuration with outer radius $h$, inner radius $h_c$, mass $M$, and spin parameter $a=J/M$, we define
$\Sexc=\pi[h_c^2-(r_-^{\Kerr})^2]$, where $r_-^{\Kerr}=M-\sqrt{M^2-a^2}$.  Under the stated regularity and interior-profile hypotheses, we prove $r_-^{\Kerr}<h_c<h$, hence $\Sexc>0$, and derive the sharp geometric bound $\Sexc<4\pi M\sqrt{M^2-a^2}$.  The Schwarzschild limit gives $\Sexc=\pi h_c^2$, whereas the Kerr endpoint gives $\Sexc=0$.  We also show that $S_+^{\Kerr}+S_-^{\Kerr}=4\pi M^2$, so that rotation at fixed mass redistributes the Kerr two-horizon reference entropy while $\Sexc$ isolates the additional internal geometric contribution.  For the time-dependent model we derive exact evolution equations for the inner root and for $\Sexc$, emphasizing that these are horizon-root bookkeeping identities unless the tracked roots are shown to be true dynamical horizons.  In the minimal $N=1$ trajectory the apparent pole at $n=2$ is removable in the mass profile, yielding a logarithmic continuation of the root equation, {although the corresponding stationary curvature is singular there.}  The stationary-profile {Ricci-scalar density is locally absolutely integrable for $-1<n\leq2$}, with a separately treated Kerr endpoint at $n=-1$. Minkowski breaking occurs at $n=0$, independently of inner-horizon disappearance.  Quasi-extremal regular configurations can approach, but not attain, the entropy bound unless the limiting geometry ceases to possess a regular Killing horizon.
\end{abstract}

\maketitle

\section{Introduction}
\label{sec:introduction}

Black-hole thermodynamics relates horizon geometry to the laws of
thermodynamics through the Bekenstein--Hawking entropy and the mechanical
first law~\cite{Bekenstein:1973ur,Hawking:1975vcx,Bardeen:1973gs}. In
Einstein gravity the entropy of a stationary bifurcate Killing horizon is
one quarter of its area, and the same result follows from the Wald
Noether-charge construction~\cite{Wald:1993nt,Iyer:1994ys,Jacobson:1993xs}.
For rotating black holes this thermodynamic structure is especially rich:
the first law contains the angular-momentum work term, while the maximally
extended Kerr geometry possesses both an outer event horizon and an inner
Cauchy horizon~\cite{Kerr:1963ud,Carter:1968rr,Teukolsky:2014vca}.

Inner horizons require substantially more care than outer horizons. Thermodynamic-like
variables associated with the Kerr inner horizon have a long history
~\cite{Curir:1979,CurirFrancaviglia:1979,OkamotoKaburaki:1992}, while modern
first-law-type relations and entropy products have been developed for a broad class
of multi-horizon geometries~\cite{Castro:2012av,Castro:2013pqa,Cvetic:2010mn,Chen:2012mh}.
At the same time, an inner Cauchy horizon is not an ordinary equilibrium
boundary accessible to an asymptotic observer, and its dynamical stability is a
separate issue. Classical mass-inflation analyses demonstrate strong curvature
growth at Cauchy horizons~\cite{Poisson:1989zz,Ori:1991zz}; for Kerr specifically,
nonlinear asymmetric perturbations produce a weak null singular behavior
~\cite{Ori:1992}, and the blue-shift instability has been established for
linearized gravitational perturbations~\cite{Sbierski:2023}. Rigorous nonlinear
results further show that a perturbed Kerr interior may remain $C^0$-extendible
across part of the Cauchy horizon even while the horizon is singular in stronger
regularity classes~\cite{DafermosLuk:2025}. Thus an inner-horizon area may be used
as a well-defined stationary geometric quantity without assuming that it has the
same operational thermodynamic meaning as the entropy of the outer event horizon.

This distinction is particularly important for regular black holes. A broad
literature has developed {regular black-hole geometries and singular geometries with locally integrable Ricci-scalar densities},
together with critical analyses of their physical viability, as possible ways
of modifying the deep interior while leaving the asymptotic black-hole charges
unchanged~\cite{Dymnikova:1992,Hayward:2006,Bambi:2013ufa,Casadio:2023iqt,
Lukash:2013ts,Ovalle:2023dtn,Arrechea:2025vcc}. Recent exact constructions and their
dynamical extensions provide Schwarzschild and Kerr-like families whose
nontrivial interiors are matched smoothly to the corresponding classical
exterior geometries and carry no additional primary conserved
charges~\cite{Ovalle:2024vuk,Ovalle:2025dyn,Ovalle:2026schw,
Ovalle:2026hair}. In the rotating family the event horizon is
exactly the Kerr event horizon for the same mass and angular momentum, while
the inner horizon depends on discrete parameters characterizing the
interior geometry~\cite{Ovalle:2026hair}.

The Schwarzschild member of this programme has recently motivated a
thermodynamic interpretation in which the regular inner Killing horizon
carries a formal geometric entropy that is absent from the final
Schwarzschild geometry~\cite{Lobo:2026entropy}. The associated area then
measures hidden internal entropy, and the discrete stationary interior
labels generate a discrete set of equilibrium geometric entropy values. A complementary
analysis compared this horizon entropy with local de Sitter
thermodynamics~\cite{Lobo:2026dS}. Those two entropy notions are distinct:
one is attached to a Killing horizon, while the other is a bulk de Sitter
construction.

Rotation changes the problem qualitatively. The terminal Kerr geometry
already possesses an inner Cauchy horizon. Consequently, the entire entropy
of a regular rotating inner horizon cannot be interpreted as entropy that
is removed during Kerr formation. The physically relevant geometric
quantity must instead measure the part of the inner-horizon area that is in
\emph{excess} of the Kerr value at the same asymptotic charges. This
observation motivates a charge-matched subtraction rather than a direct
transplantation of the Schwarzschild formula.

The construction studied here is based on the regular Kerr-like family of
Ref.~\cite{Ovalle:2026hair} and on its time-dependent extension describing
the final stage of Kerr formation~\cite{Ovalle:2026wqw}. The latter
keeps the ADM mass fixed while promoting the rotation parameter and the
interior profile to time-dependent quantities. Importantly, the evolving
roots of the Kerr-like function \(\Delta(v,r)\) have not yet been proven to
coincide with true marginal or dynamical horizons. The distinction between
roots, trapping horizons, isolated horizons, and dynamical horizons is
essential in nonstationary gravity~\cite{Hayward:1993wb,Ashtekar:2001is,
Ashtekar:2003hk,AshtekarKrishnan:2004,Booth:2005qc,Ashtekar:2025far}. We therefore separate the
stationary horizon geometry and area-law bookkeeping, which are exact, from
the dynamical \emph{horizon-root entropy diagnostic}, which is an algebraically exact
bookkeeping construction but not yet a dynamical-horizon first law.

Our main results are as follows. We define the Kerr-subtracted
inner-horizon entropy and prove that it is positive throughout the regular
nonextremal family under the interior mass-profile assumptions stated below. We derive a family-wide upper bound and show that the bound is a
supremum: its strict saturation at fixed nonextremal spin corresponds to a
Kerr mimicker with a cusp at the nominal horizon rather than a regular
Killing horizon. We relate the excess entropy to the deformation of the
Kerr entropy product and explain why this does not conflict with known
area-product relations, whose hypotheses are different. We then derive the
exact evolution equation of the inner root, regularize the apparent
\(n=2\) pole in the minimal collapse profile, and formulate a conditional
two-horizon geometric bookkeeping entropy balance that is deliberately distinguished from the
standard generalized second law (GSL) for causal horizons~\cite{Gao:2001ut,Wall:2011hj}. The stationary-profile analysis in Appendix~\ref{app:integrability-MB} distinguishes {stationary-profile curvature singularity onset at $n=2$}, Minkowski breaking at $n=0$, and the Kerr endpoint at $n=-1$, without identifying Minkowski breaking with inner-horizon disappearance.

The paper is organized as follows. Section~\ref{sec:geometry} reviews the
stationary geometry and its horizon data. Section~\ref{sec:excess}
introduces the excess entropy, proves its positivity and bound, and relates
it to entropy products. Section~\ref{sec:thermodynamics} develops the
physical interpretation of the stationary construction. The exact
horizon-root dynamics and the regularized \(N=1\) trajectory are analysed in
Sec.~\ref{sec:dynamics}. Section~\ref{sec:balance_capacity} develops the conditional
two-horizon geometric bookkeeping entropy balance and analyses the
quasi-extremal entropy capacity of the regular rotating family. The relation
to the previous Schwarzschild and de Sitter constructions, together with the
scope, limitations, and principal open problems of the framework, is discussed
in Sec.~\ref{sec:relation_scope}, and we conclude in
Sec.~\ref{sec:conclusions}.

\section{Regular rotating geometry and stationary horizon data}
\label{sec:geometry}

\subsection{Piecewise Kerr-like geometry}

We work in geometrized units \(G=c=\hbar=k_{\rm B}=1\). The stationary
geometry is of G\"urses--G\"ursey Kerr--Schild form~\cite{Gurses:1975vu},
\begin{align}
 ds^2={}&-\left(1-\frac{2r\mtilde(r)}{\rho^2}\right)dt^2
 -\frac{4ar\mtilde(r)\sin^2\theta}{\rho^2}\,dt\,d\phi
 \notag\\
 &+\frac{\rho^2}{\Delta}\,dr^2+\rho^2d\theta^2
 +\frac{\Sigma\sin^2\theta}{\rho^2}\,d\phi^2,
 \label{eq:stationary-metric}
\end{align}
where
\begin{align}
 \rho^2&=r^2+a^2\cos^2\theta,
 \label{eq:rho}\\
 \Delta(r)&=r^2-2r\mtilde(r)+a^2,
 \label{eq:Delta}\\
 \Sigma&=(r^2+a^2)^2-a^2\Delta\sin^2\theta,
 \label{eq:Sigma}
\end{align}
and $ a=J/M$.

The full mass profile is piecewise,
\begin{equation}
 \mtilde(r)=
 \begin{cases}
 m(r),&0<r\leq h,\\
 M,&r>h,
 \end{cases}
 \label{eq:piecewise-static}
\end{equation}
so the stationary exterior is exactly Kerr. The function \(m(r)\) denotes
the interior mass profile inherited from the corresponding spherical seed;
it should not be interpreted as a generic axisymmetric Misner--Sharp mass.

Smooth matching at the outer horizon requires
$ m(h)=M$, and $m'(h)=0$,
with
\begin{equation}
 h=M+\sqrt{M^2-a^2}.
 \label{eq:outer-radius}
\end{equation}
For the regular rotating branch we take \(0<|a|<M\) and the stationary
interior exponents satisfy \(n_i>2\). The explicit family has a monotone
interior mass profile, in particular
$ 0<m(r)<M$ and $0<r<h$, with \(m'(r)>0\) on the interior branch~\cite{Ovalle:2026hair,Ovalle:2026wqw}.

Besides the outer zero \(r=h\), the stationary regular solutions possess a
single positive inner zero
\begin{equation}
 \Delta(h_c)=0,
 \qquad
 0<h_c<h,
 \label{eq:hc}
\end{equation}
which is the stationary inner Cauchy horizon of the geometry. The Kerr
radii associated with the same \(M\) and \(a\) are
\begin{equation}
 r_\pm^{\Kerr}=M\pm\sqrt{M^2-a^2},
 \label{eq:Kerr-radii}
\end{equation}
so that, by construction, $ h=r_+^{\Kerr}$.

The regular interior is supported by an anisotropic effective source. For
the stationary Kerr--Schild family its principal components can be written
schematically as~\cite{Ovalle:2026hair}
\begin{align}
 8\pi\epsilon&=-8\pi p_r=\frac{2r^2}{\rho^4}m'(r),
 \label{eq:source1}\\
 8\pi p_\theta=8\pi p_\phi
 &=-\frac{r}{\rho^2}m''(r)
 +\frac{2(r^2-\rho^2)}{\rho^4}m'(r).
 \label{eq:source2}
\end{align}
The nonvacuum interior will be relevant below when comparing our entropy
product with electrovacuum area-product theorems.

\subsection{Horizon area, angular velocity, and surface gravity}

Let \(r=r_H\) be a regular stationary root of \(\Delta(r)=0\). On a spatial
cross-section of the corresponding Killing horizon,
$ \Sigma_H=(r_H^2+a^2)^2$.
The induced two-metric therefore gives
$ \sqrt{g_{\theta\theta}g_{\phi\phi}}
 =(r_H^2+a^2)\sin\theta$,
and the area is
\begin{equation}
 A_H=4\pi(r_H^2+a^2).
 \label{eq:horizon-area}
\end{equation}
In Einstein gravity the corresponding stationary geometric entropy is
\begin{equation}
 S_H=\frac{A_H}{4}
 =\pi(r_H^2+a^2).
 \label{eq:horizon-entropy}
\end{equation}
Thus
\begin{align}
 S_+&=\pi(h^2+a^2),
 \label{eq:Splus}\\
 S_c&=\pi(h_c^2+a^2).
 \label{eq:Sc}
\end{align}

The horizon angular velocity has the Kerr-like form
\begin{equation}
 \Omega_H=\frac{a}{r_H^2+a^2},
 \label{eq:OmegaH}
\end{equation}
and, for a nondegenerate stationary root, the signed surface gravity is
\begin{equation}
 \kappa_H
 =\frac{\Delta'(r_H)}{2(r_H^2+a^2)}
 =\frac{r_H-m(r_H)-r_Hm'(r_H)}{r_H^2+a^2}.
 \label{eq:kappa-general}
\end{equation}
At the outer horizon the matching conditions give
\begin{equation}
 \kappa_+
 =\frac{\sqrt{M^2-a^2}}{h^2+a^2}>0,
 \label{eq:kappa-plus}
\end{equation}
which is exactly the Kerr value and is independent of the interior labels.
For a simple inner horizon with the standard root ordering,
$ \kappa_c<0$.
When useful, a positive magnitude may be introduced,
$ T_c^{\rm geom}\equiv |\kappa_c|/(2\pi)$,
but we shall regard this only as a stationary geometric temperature scale,
not as the temperature of an ordinary thermal bath seen from infinity.

\section{Kerr-subtracted inner-horizon entropy}
\label{sec:excess}

\subsection{The Kerr two-horizon reference state}

For Kerr,
\begin{equation}
 S_\pm^{\Kerr}
 =\pi\left[(r_\pm^{\Kerr})^2+a^2\right].
 \label{eq:SKpm}
\end{equation}
Using $ r_+^{\Kerr}+r_-^{\Kerr}=2M$, and $r_+^{\Kerr}r_-^{\Kerr}=a^2$,
one obtains
\begin{equation}
 S_\pm^{\Kerr}=2\pi Mr_\pm^{\Kerr}.
 \label{eq:SKpm-simple}
\end{equation}
The sum is therefore
\begin{equation}
 S_+^{\Kerr}+S_-^{\Kerr}=4\pi M^2,
 \label{eq:Kerr-sum}
\end{equation}
while the product is
\begin{equation}
 S_+^{\Kerr}S_-^{\Kerr}=4\pi^2J^2.
 \label{eq:Kerr-product}
\end{equation}
Equation~\eqref{eq:Kerr-sum} is a two-horizon geometric identity. It is
not the thermodynamic entropy measured by a single asymptotic observer.
The product~\eqref{eq:Kerr-product} belongs to the broader class of
mass-independent horizon products that motivated a substantial literature
on inner-horizon mechanics~\cite{Cvetic:2010mn,Castro:2012av,Castro:2013pqa}.
Historically, the attribution of thermodynamic-like variables to the Kerr inner
horizon predates this modern area-product literature
~\cite{Curir:1979,CurirFrancaviglia:1979,OkamotoKaburaki:1992}. In the present
work the two-horizon Kerr quantities are used only as a charge-matched geometric
reference; the new object is the excess relative to that reference, not the
assignment $A_-^{\Kerr}/4$ itself.

\subsection{Definition and exact decomposition}

The outer horizon of the regular solution coincides geometrically with the
Kerr outer horizon for the same charges,
$ S_+=S_+^{\Kerr}$.
The departure from Kerr in a two-horizon bookkeeping therefore resides
entirely in the inner horizon. We define the \emph{excess inner-horizon
entropy}
\begin{equation}
 \Sexc\equiv S_c-S_-^{\Kerr}.
 \label{eq:Sexc-def}
\end{equation}
The explicit \(a^2\) terms cancel, giving
\begin{equation}
 \Sexc
 =\pi\left[h_c^2-(r_-^{\Kerr})^2\right].
 \label{eq:Sexc}
\end{equation}
Thus rotation affects both horizon areas, but the extra contribution beyond
Kerr is controlled solely by the radial displacement of the regular inner
horizon relative to the Kerr Cauchy horizon.

Define the regular two-horizon geometric bookkeeping entropy by
$ \Sgeomreg\equiv S_++S_c $.
Then
\begin{equation}
 \Sgeomreg=4\pi M^2+\Sexc.
 \label{eq:Sgeom-decomp}
\end{equation}
Correspondingly,
\begin{equation}
 \SgeomKerr\equiv S_+^{\Kerr}+S_-^{\Kerr}=4\pi M^2,
 \label{eq:Sgeom-K}
\end{equation}
so that
\begin{equation}
 \Sgeomreg-\SgeomKerr=\Sexc.
 \label{eq:Sgeom-difference}
\end{equation}
This identity is the precise sense in which \(\Sexc\) measures the
additional stationary internal geometry relative to a charge-matched Kerr
reference state.

\subsection{Positivity theorem}

\begin{theorem}[Positivity of the Kerr-subtracted inner entropy]
\label{thm:positive}
Consider a nonextremal stationary rotating configuration with
\(0<|a|<M\), outer radius \(h=r_+^{\Kerr}\), and an interior mass profile
satisfying $m\in C^1(0,h]$ and
\begin{align}
 m(h)&=M,
 &m'(h)&=0,
 \notag\\
 m(r)&<M,
 &0&<r<h.
 \label{eq:theorem-assumptions}
\end{align}
Assume that \(\Delta(r)\) possesses a single regular inner root \(h_c\) in
\(0<r<h\). Then
\begin{equation}
 r_-^{\Kerr}<h_c<h,
 \label{eq:hc-bound}
\end{equation}
and hence
$ \Sexc>0$.
\end{theorem}

\begin{proof}
On the interior branch,
\begin{equation}
 \Delta(r)=\Delta_{\Kerr}(r)+2r[M-m(r)],
 \label{eq:Delta-difference}
\end{equation}
where $ \Delta_{\Kerr}(r)=r^2-2Mr+a^2$.
Since \(r_-^{\Kerr}>0\) and
\(\Delta_{\Kerr}(r_-^{\Kerr})=0\),
\begin{equation}
 \Delta(r_-^{\Kerr})
 =2r_-^{\Kerr}[M-m(r_-^{\Kerr})]>0.
 \label{eq:Delta-rminus-positive}
\end{equation}
At the common outer root,
\begin{align}
 \Delta'(h)
 &=2h-2M-2hm'(h)
 \notag\\
 &=2\sqrt{M^2-a^2}>0.
 \label{eq:Delta-prime-h}
\end{align}
Therefore \(\Delta(r)<0\) immediately inside \(h\). The assumed continuity of $m$, and hence of $\Delta$, forces at
least one zero in \((r_-^{\Kerr},h)\). Under the single-inner-root
hypothesis this zero is \(h_c\), proving Eq.~\eqref{eq:hc-bound} and hence
\(\Sexc>0\).
\end{proof}

The Schwarzschild limit \(a=0\), where \(r_-^{\Kerr}=0\), is treated
separately; in that case \(\Sexc=\pi h_c^2>0\) follows directly from
\(h_c>0\).

\subsection{Geometric upper bound and its status as a supremum}

Since \(h_c<h=r_+^{\Kerr}\), Theorem~\ref{thm:positive} gives
\begin{equation}
 0<\Sexc
 <\pi\left[(r_+^{\Kerr})^2-(r_-^{\Kerr})^2\right].
\end{equation}
Therefore
\begin{equation}
 0<\Sexc<4\pi M\sqrt{M^2-a^2}.
 \label{eq:Sexc-bound}
\end{equation}
Introducing
\begin{equation}
 \chi\equiv\frac{a}{M},
 \qquad
 \etaexc\equiv\frac{\Sexc}{4\pi M^2},
 \label{eq:chi-eta}
\end{equation}
we obtain
\begin{equation}
 0<\etaexc<\sqrt{1-\chi^2}.
 \label{eq:eta-bound}
\end{equation}
At fixed mass the available excess-entropy capacity therefore shrinks as
the true Kerr extremal limit \(|\chi|\to1\) is approached.

The upper value in Eq.~\eqref{eq:Sexc-bound} is a \emph{supremum} of the
regular nonextremal family, not a maximum attained by a regular black hole.
For large interior exponents the stationary solutions satisfy
\(h_c\to h\), but the strict limiting profile
\begin{equation}
 m(r)\longrightarrow M\left(\frac{r}{h}\right)^3
 \label{eq:mimicker-profile}
\end{equation}
produces a Kerr mimicker: the surface \(r=h\) remains a null
infinite-redshift surface but \(\Delta\) develops a cusp, so the limiting
surface is not a regular Killing horizon~\cite{Ovalle:2026hair}. Thus
regular quasi-extremal configurations can approach the entropy bound
arbitrarily closely while strict saturation lies outside the regular
black-hole class.

\subsection{Schwarzschild limit and Kerr endpoint}

In the nonrotating limit, $ a\to0$, $r_-^{\Kerr}\to 0$, and $ h\to 2M$, so $ \Sexc\to\pi h_c^2$.
This is exactly the hidden inner-horizon entropy used in the regular Schwarzschild construction~\cite{Lobo:2026entropy}.

At the Kerr endpoint of the interpolating trajectory,
$ m(r)\to M$, $ h_c\to r_-^{\Kerr}$, and hence
$ \Sexc\to 0$.
The statement refers to the limiting value of the entropy diagnostic. A
continuous parameter trajectory from a regular state to Kerr need not
remain within the regular sector; indeed, the explicit collapse trajectory
crosses {a singular stationary-profile regime with a locally integrable Ricci-scalar density} before reaching \(n=-1\), as discussed
in Sec.~\ref{sec:n2}.

\subsection{Entropy-product deformation and comparison with area-product results}

For the regular geometry,
\begin{align}
 S_+S_c
 &=S_+\left(S_-^{\Kerr}+\Sexc\right)
 \notag\\
 &=4\pi^2J^2+S_+\Sexc.
 \label{eq:product-deformation}
\end{align}
Therefore
\begin{equation}
 S_+S_c-4\pi^2J^2=S_+\Sexc.
 \label{eq:product-excess}
\end{equation}
The same internal geometric displacement that defines \(\Sexc\) therefore
measures the departure from the Kerr entropy product.

This result should not be interpreted as contradicting universal
area-product theorems. For stationary axisymmetric black holes with an
electrovacuum region satisfying the hypotheses of Ansorg and Hennig, one
finds, for example, \(A_+A_-=(8\pi J)^2\) in the uncharged case and its
Einstein--Maxwell extension in the charged case~\cite{Ansorg:2009yi}.
The present regular interior is instead supported by the nonvacuum
anisotropic source in Eqs.~\eqref{eq:source1}--\eqref{eq:source2}; the
relevant electrovacuum hypotheses are therefore not satisfied. More
generally, mass-independent area products are not universal once the matter
content or gravitational dynamics is changed~\cite{Castro:2013pqa,Visser:2012wu,Pradhan:2016}.
Equation~\eqref{eq:product-excess} is consequently best viewed as an exact
measure of the matter-supported deformation away from the Kerr product.

\section{Stationary thermodynamic interpretation}
\label{sec:thermodynamics}

\subsection{What is, and is not, being called entropy}

For each regular stationary Killing horizon we adopt the standard
Einstein-gravity Noether-charge assignment \(S_H=A_H/4\), with the usual
stationary-horizon interpretation~\cite{Wald:1993nt,Iyer:1994ys,
Jacobson:1993xs}. For the inner
horizon, however, we do not assume an ordinary canonical ensemble or a
thermal bath accessible to an external observer. The quantity \(S_c\) is
used as a formal stationary geometric entropy in the same restricted sense
as in the historical and modern inner-horizon mechanics literature
~\cite{Curir:1979,OkamotoKaburaki:1992,Castro:2012av,Chen:2012mh}.

The subtraction in Eq.~\eqref{eq:Sexc-def} is then physically motivated by
the endpoint geometry. The regular inner horizon does not disappear in the
rotating transition; it approaches the Kerr Cauchy horizon. Thus the
rotational process is
\begin{equation}
 \text{regular inner horizon}
 \longrightarrow
 \text{Kerr inner horizon},
 \label{eq:rot-process}
\end{equation}
whereas the corresponding Schwarzschild process ends with no inner horizon.
Only the difference \(S_c-S_-^{\Kerr}\) is absent from the final Kerr
reference state.

\subsection{Hidden interior state information at fixed asymptotic charges}

A notable feature of the stationary family is that different sets of
interior integers \(\{n_i\}\) can share the same asymptotic charges
\((M,J)\) and the same Kerr exterior, while possessing different values of
\(h_c\). Consequently,
\begin{equation}
 \Sexc=\Sexc(M,J;\{n_i\})
 \label{eq:Sexc-state}
\end{equation}
varies among geometries that are indistinguishable by the primary charges
alone. The quantity \(\Sexc\) therefore provides a thermodynamic diagnostic
of \emph{secondary interior structure}: information stored behind the outer
horizon without introducing a new conserved charge at infinity.

For the stationary regular family the exponents satisfy
\(n_i\in\mathbb N\), \(n_i>2\). At fixed \(M\), \(J\), and \(N\), the allowed
inner roots form a discrete set,
\begin{equation}
 h_c=h_c(M,J;\{n_i\}),
\end{equation}
and hence
\begin{equation}
 \Sexc(M,J;\{n_i\})
 =\pi\left[h_c^2(M,J;\{n_i\})-(r_-^{\Kerr})^2\right].
 \label{eq:discrete-spectrum}
\end{equation}
This is a discrete, possibly degenerate, set of geometric entropy values associated with equilibrium configurations. For fixed $N$, large-integer sequences may accumulate
toward the quasi-extremal boundary of the regular family, where the upper entropy
bound is approached but not attained by a regular Killing horizon. This discrete set should not yet be identified with a microscopic or quantum spectrum. In particular,
the continuous functions \(n_i(v)\) introduced in the collapse model are
interpolation variables and need not remain integer away from stationary endpoints.

\subsection{Outer first law and Kerr entropy redistribution}

Because the stationary exterior and outer horizon are exactly Kerr, $ h^2+a^2=2Mh$ and $ S_+=2\pi Mh$.
With
\begin{equation}
 T_+=\frac{\kappa_+}{2\pi},
 \qquad
 \Omega_+=\frac{a}{h^2+a^2},
\end{equation}
the standard first law holds,
\begin{equation}
 dM=T_+\,dS_++\Omega_+\,dJ.
 \label{eq:first-law-outer}
\end{equation}
The isolated-horizon framework provides a quasi-local setting in which
rotating horizon mechanics can be formulated without requiring a globally
stationary exterior~\cite{Ashtekar:2001is}; nevertheless, in the present
stationary sector Eq.~\eqref{eq:first-law-outer} is simply the Kerr law.

At fixed $M$, write the spin magnitude as $j\equiv |a|$ and define
\begin{equation}
 \delta\equiv\sqrt{M^2-j^2}.
\end{equation}
Then
\begin{equation}
 S_\pm^{\Kerr}=2\pi M(M\pm\delta),
\end{equation}
and therefore
\begin{equation}
 \frac{dS_+^{\Kerr}}{dj}
 =-\frac{2\pi Mj}{\delta},
 \qquad
 \frac{dS_-^{\Kerr}}{dj}
 =+\frac{2\pi Mj}{\delta}.
 \label{eq:Kerr-redistribution-derivatives}
\end{equation}
Thus increasing the spin magnitude $|a|=j$ at fixed mass decreases the outer Kerr entropy and increases the inner Kerr entropy by exactly the same amount. Equation~\eqref{eq:Kerr-sum}
is not merely an algebraic curiosity: it cleanly separates a \emph{Kerr
redistribution sector}, whose two-horizon geometric bookkeeping sum is fixed, from the additional
regular-interior contribution \(\Sexc\).

\subsection{Inner mechanics and the absence of an assumed interior first law}

For a stationary regular inner horizon, the signed surface gravity is given
by Eq.~\eqref{eq:kappa-general}. Inner-horizon first laws are known for many
standard stationary black holes~\cite{Castro:2012av,Chen:2012mh}, but we do
not assume a two-variable Kerr-like first law for the regular inner horizon.
The labels \(\{n_i\}\) distinguish interior geometries while carrying no
primary asymptotic charge, and they are discrete in the regular stationary
sector. A Hamiltonian or covariant-phase-space derivation would be required
to determine whether changes of interior structure admit conjugate
potentials or whether only transitions between distinct stationary states
are meaningful.

This caution is important. The identity \(S_c=A_c/4\) is a stationary
Noether-charge statement for a formal geometric entropy; it does not by itself imply a thermodynamic first
law involving arbitrary variations of the interior labels. The present
framework therefore uses \(S_c\) and \(\Sexc\) as geometric state functions
without postulating additional work terms.

\section{Time-dependent Kerr formation and horizon-root entropy}
\label{sec:dynamics}

\subsection{Evolving geometry and status of the roots}

The time-dependent construction promotes
\begin{equation}
 a\to a(v),
 \qquad
 \mtilde(r)\to\mtilde(v,r),
\end{equation}
with the piecewise profile
\begin{equation}
 \mtilde(v,r)=
 \begin{cases}
 m(v,r),&0<r\leq h(v),\\
 M,&{r>h(v)},
 \end{cases}
 \label{eq:piecewise-dyn}
\end{equation}
and
\begin{equation}
 \Delta(v,r)=r^2-2r\mtilde(v,r)+a^2(v).
 \label{eq:Delta-dynamic}
\end{equation}
The outer root is prescribed by
\begin{equation}
 h(v)=M+\sqrt{M^2-a^2(v)},
 \label{eq:h-dynamic}
\end{equation}
with fixed ADM mass \(M\). Throughout the dynamical discussion we assume $|a(v)|<M$, whenever expressions containing $(M^2-a^2)^{-1/2}$ are used. 

If the prescribed
final state is extremal, $|a_f|=M$, all such formulas are understood in the
asymptotic nonextremal limit $v\to+\infty$. The interior equation possesses a second
positive root \(h_c(v)\),
\begin{equation}
 \Delta(v,h_c(v))=0.
 \label{eq:hc-dynamic}
\end{equation}
All formulas below that involve $h_c(v)$ are valid on intervals where the inner branch being followed exists continuously, is unique within that branch, and remains nondegenerate. If an additional positive root appears, or if $\partial_r\Delta$ vanishes, the algebraic root-tracking description must be restarted with a separate branch analysis. The collapse construction tracks these two roots throughout the evolution~\cite{Ovalle:2026wqw}.

In a nonstationary axisymmetric spacetime, however, the equation
\(\Delta(v,r)=0\) alone does not establish that the corresponding surface
is an event, apparent, trapping, isolated, or dynamical horizon. A genuine
quasi-local horizon is defined through the null expansions of appropriate
two-surfaces, and a physical dynamical first law requires the associated
matter and gravitational fluxes~\cite{Hayward:1993wb,Ashtekar:2003hk,
Booth:2005qc,Ashtekar:2025far}. We therefore call \(h(v)\) and \(h_c(v)\)
\emph{evolving horizon roots}. All time-dependent entropy expressions below
are exact functions of these roots, but they are not claimed to be the
entropy of a rigorously identified dynamical horizon.

\subsection{Outer-root mechanical identity}

Differentiating Eq.~\eqref{eq:h-dynamic} gives
\begin{equation}
 \dot h=-\frac{a\dot a}{\sqrt{M^2-a^2}}.
 \label{eq:hdot}
\end{equation}
Define the instantaneous Kerr expressions
\begin{align}
 S_+(v)&=\pi[h^2(v)+a^2(v)],\\
 \Omega_+(v)&=\frac{a(v)}{h^2(v)+a^2(v)},\\
 T_+(v)&=\frac{1}{2\pi}
 \frac{\sqrt{M^2-a^2(v)}}{h^2(v)+a^2(v)}.
\end{align}
Since \(J(v)=Ma(v)\), one obtains identically
\begin{equation}
 T_+\dot S_++\Omega_+\dot J=0
 \qquad(M=\mathrm{const.}).
 \label{eq:dynamic-outer-law}
\end{equation}
This is the formal first-law-type identity of the collapse construction~\cite{Ovalle:2026wqw}. It is an exact consequence of the
Kerr-like outer-root relation; interpreting it as a physical dynamical
first law requires identifying the relevant dynamical horizon and fluxes.

\subsection{Exact transport equation for the inner root}

Implicit differentiation of Eq.~\eqref{eq:hc-dynamic} gives
\begin{equation}
 0=\left.\partial_v\Delta\right|_{h_c}
 +\left.\partial_r\Delta\right|_{h_c}\dot h_c.
\end{equation}
On the interior branch,
\begin{equation}
 \partial_v\Delta=-2r\,m_{,v}+2a\dot a,
 \qquad
 \partial_r\Delta=2r-2m-2rm_{,r}.
\end{equation}
Along such a continuous nondegenerate branch,
\(\partial_r\Delta|_{h_c}\neq0\), and hence
\begin{equation}
 \dot h_c
 =\frac{h_c\,m_{,v}(v,h_c)-a\dot a}
 {h_c-m(v,h_c)-h_c m_{,r}(v,h_c)}.
 \label{eq:hcdot}
\end{equation}
This equation is exact and separates the root motion generated by the
evolving interior mass profile from the explicit rotational contribution. It ceases to be valid at a degenerate root, where $\partial_r\Delta|_{h_c}=0$.

\subsection{The minimal \texorpdfstring{\(N=1\)}{N=1} trajectory and the removable \texorpdfstring{\(n=2\)}{n=2} pole}
\label{sec:n2}

For the minimal model,
\begin{equation}
 m(v,r)=M\,\mu(x,n),
 \qquad
 x\equiv\frac{r}{h(v)},
 \label{eq:m-mu}
\end{equation}
where, away from \(n=2\),
\begin{equation}
 \mu(x,n)=\frac{(n+1)x^3-3x^{n+1}}{n-2}.
 \label{eq:mu-naive}
\end{equation}
The Kerr-forming trajectory, {with $0<|a_f|\leq M$}, is chosen as~\cite{Ovalle:2026wqw}
\begin{equation}
 a(v)=\frac{a_f}{2}\,[1+\tanh(\omega v)],\qquad \omega>0,
 \label{eq:a-profile}
\end{equation}
with
\begin{equation}
 n(v)=\alpha-\frac{1+\alpha}{a_f}\,a(v),
 \qquad
 \alpha>2,
 \label{eq:n-profile}
\end{equation}
so that
\begin{equation}
 n(-\infty)=\alpha,
 \qquad
 n(+\infty)=-1.
\end{equation}
Every such trajectory necessarily crosses \(n=2\). The apparent pole in
Eq.~\eqref{eq:mu-naive} is nevertheless removable at the level of the mass
profile.

\begin{proposition}[Continuous mass-profile extension through \(n=2\)]
For \(0<x\leq1\), the function \(\mu(x,n)\) has the finite limit
\begin{equation}
 \lim_{n\to2}\mu(x,n)=x^3\left(1-3\ln x\right).
 \label{eq:mu-limit}
\end{equation}
Accordingly, the continuous definition is
\begin{equation}
 \mu(x,n)=
 \begin{cases}
 \dfrac{(n+1)x^3-3x^{n+1}}{n-2},&n\neq2,\\[8pt]
 x^3(1-3\ln x),&n=2.
 \end{cases}
 \label{eq:mu-continued}
\end{equation}
\end{proposition}

\begin{proof}
Both numerator and denominator of Eq.~\eqref{eq:mu-naive} vanish at
\(n=2\). L'H\^opital's rule gives
\begin{equation}
 \lim_{n\to2}\mu(x,n)
 =x^3-3x^3\ln x,
\end{equation}
which is Eq.~\eqref{eq:mu-limit}.
\end{proof}

The continuation preserves the matching conditions at the outer root:
\begin{equation}
 \mu(1,2)=1,
 \qquad
 \left.\partial_x\mu(x,2)\right|_{x=1}=0,
\end{equation}
with
\begin{equation}
 \partial_x\mu(x,2)=-9x^2\ln x.
 \label{eq:mux-n2}
\end{equation}
For numerical evolution it is also useful that
\begin{equation}
 \left.\partial_n\mu(x,n)\right|_{n=2}
 =-\frac{3}{2}x^3(\ln x)^2.
 \label{eq:mun-n2}
\end{equation}
At fixed coordinate radius \(r\), one may therefore evaluate
\begin{equation}
 m_{,v}
 =M\left[-x\frac{\dot h}{h}\,\mu_{,x}
 +\dot n\,\mu_{,n}\right]
 \label{eq:mv-mu}
\end{equation}
continuously through \(n=2\).

The removability of the pole in \(m\) does \emph{not} mean that the geometry
remains regular. The stationary analysis of the same family shows that the
Kretschmann scalar develops logarithmic singular behavior at \(n=2\).
The singular-sector interval relevant to this interpolation is $-1\leq n\leq2$. More precisely, the stationary-profile
{Ricci-scalar density} is locally absolutely integrable for $-1<n\leq2$, while the endpoint
$n=-1$ is Ricci-flat away from the Kerr ring singularity; Appendix~\ref{app:integrability-MB}
treats these endpoints explicitly. This test does not establish integrability of
{other curvature invariants, including the Kretschmann scalar, or regularity of the full time-dependent geometry.}
\cite{Ovalle:2026hair,Lukash:2013ts,Ovalle:2023dtn,Arrechea:2025vcc}
Thus the root equation and the entropy diagnostic can be
continued through \(n=2\), but the Kerr-forming trajectory has already left
the regular-black-hole sector. {The threshold $n=2$ concerns the stationary
profiles; time dependence of $a(v)$ already induces curvature singularities
in the full evolving geometry~\cite{Ovalle:2026wqw}.}

The crossing occurs at
\begin{equation}
 a_*=a_f\frac{\alpha-2}{\alpha+1},
 \label{eq:a-star}
\end{equation}
and, for the hyperbolic-tangent profile,
\begin{equation}
 \tanh(\omega v_*)=\frac{\alpha-5}{\alpha+1}.
 \label{eq:v-star}
\end{equation}
The Kerr state is reached only asymptotically at \(n=-1\).
Minkowski breaking, however, already occurs at $n=0$ in the
stationary profiles: the equatorial core limit of $g_{tt}$ changes from $-1$
for $n>0$ to $-1+3M/h$ at $n=0$, as shown in Eq.~\eqref{MBx}.
For the prescribed interpolation this parameter value is attained at
$a_{\rm MB}=a_f\alpha/(\alpha+1)$ and
$v_{\rm MB}=\ln\alpha/(2\omega)$, after the $n=2$ crossing and before the
asymptotic Kerr endpoint. The core-limit change does not require the inner
root to disappear. {Stationary-profile curvature singularity onset}, Minkowski breaking, and the Kerr endpoint
are therefore distinct stages of this trajectory, rather than a single
horizon-loss event.

\subsection{Dimensionless root equation}

Introduce
\begin{equation}
 \chi(v)\equiv\frac{a(v)}{M},
 \qquad
 H(\chi)\equiv\frac{h(v)}{M}
 =1+\sqrt{1-\chi^2}.
\end{equation}
The interior horizon-root equation becomes
\begin{equation}
 H^2x^2-2Hx\,\mu(x,n)+\chi^2=0.
 \label{eq:dimensionless-root}
\end{equation}
The outer solution is \(x=1\). The second positive solution \(x_c<1\),
when present, gives
\begin{equation}
 h_c(v)=h(v)x_c(v).
\end{equation}
Because Eq.~\eqref{eq:mu-continued} is finite at \(n=2\),
Eq.~\eqref{eq:dimensionless-root} supplies a continuous algebraic route for
tracking the inner root provided the chosen branch remains single and nondegenerate, even across {the stationary-profile singularity threshold}.

\subsection{Dynamical excess-entropy diagnostic}

At each \(v\), define the charge-matched instantaneous Kerr inner radius
\begin{equation}
 r_-^{\Kerr}(v)=M-\sqrt{M^2-a^2(v)}.
 \label{eq:rminus-dyn}
\end{equation}
The horizon-root excess entropy is
\begin{equation}
 \Sexc(v)=\pi\left[h_c^2(v)-\big(r_-^{\Kerr}(v)\big)^2\right].
 \label{eq:Sexc-dyn}
\end{equation}
Its derivative is exactly
\begin{equation}
 \Sexcdot
 =2\pi\left[h_c\dot h_c-r_-^{\Kerr}\dot r_-^{\Kerr}\right],
 \label{eq:Sexc-dot}
\end{equation}
where
\begin{equation}
 \dot r_-^{\Kerr}
 =\frac{a\dot a}{\sqrt{M^2-a^2}}.
 \label{eq:rminus-dot}
\end{equation}
Combining Eqs.~\eqref{eq:hcdot} and~\eqref{eq:rminus-dot} yields
\begin{align}
 \frac{\Sexcdot}{2\pi}
 ={}&
 \frac{h_c\left[h_c m_{,v}(v,h_c)-a\dot a\right]}
 {h_c-m(v,h_c)-h_c m_{,r}(v,h_c)}
 \notag\\
 &-r_-^{\Kerr}\frac{a\dot a}{\sqrt{M^2-a^2}}.
 \label{eq:Sexc-dot-explicit}
\end{align}
Equivalently, in dimensionless variables,
\begin{equation}
 \frac{\Sexc(v)}{\pi M^2}
 =H^2x_c^2-\left(1-\sqrt{1-\chi^2}\right)^2.
 \label{eq:Sexc-dimensionless}
\end{equation}

Assuming continuous tracking of the relevant inner branch to the Kerr endpoint, the endpoint identity is algebraic even without proving local
monotonicity. For a regular initial state, \(\Sexc(v_i)>0\), whereas the
Kerr endpoint satisfies \(\Sexc(v_f)=0\). Hence
\begin{equation}
 \Delta\Sexc
 \equiv\Sexc(v_f)-\Sexc(v_i)
 =-\Sexc(v_i)<0.
 \label{eq:net-excess-change}
\end{equation}
The exact transport equation~\eqref{eq:Sexc-dot-explicit} determines whether
this net decrease is monotonic or proceeds through intermediate storage and
release phases. No monotonicity theorem follows from the root identities
alone.

\section{Conditional entropy balance and quasi-extremal entropy capacity}
\label{sec:balance_capacity}

The standard generalized second law (GSL) concerns the generalized entropy
associated with a causal horizon, namely the horizon entropy together with
the entropy of quantum fields outside that horizon
\cite{Gao:2001ut,Wall:2011hj}. The two-horizon geometric bookkeeping
	sum \(S_++S_c\) used here is not the standard generalized entropy, because
it includes the area of an inner Cauchy horizon. Accordingly, no GSL is
inferred directly from \(\Sgeom\).

Instead, we introduce the explicit bookkeeping quantity
\begin{equation}
	\Sbook(v)\equiv S_{\rm env}(v)+\Sgeom(v),
	\label{eq:Sbook}
\end{equation}
where
\begin{equation}
	\Sgeom(v)\equiv S_+(v)+S_c(v)
	=4\pi M^2+\Sexc(v).
	\label{eq:Sgeom-dyn}
\end{equation}
At fixed ADM mass,
\begin{equation}
	\frac{d\Sgeom}{dv}=\Sexcdot.
	\label{eq:Sgeom-dot}
\end{equation}
This follows because the spin-induced decrease of the Kerr outer entropy is
exactly compensated by the increase of the Kerr inner reference entropy,
as shown in Eq.~\eqref{eq:Kerr-redistribution-derivatives}. The only
time-dependent contribution to the two-horizon geometric bookkeeping sum
is therefore the excess term \(\Sexc\).

A conditional entropy-balance statement can then be formulated. If the
physical process admits a coarse-grained description in which
\(\Sbook\) is nondecreasing,
\begin{equation}
	\frac{d\Sbook}{dv}\geq0,
	\label{eq:book-second-law}
\end{equation}
then
\begin{equation}
	\dot S_{\rm env}\geq-\Sexcdot.
	\label{eq:conditional-balance}
\end{equation}
Assuming that the inner-root branch can be followed continuously from a
regular initial configuration to the Kerr endpoint of the interpolating
trajectory, integration gives
\begin{equation}
	\Delta S_{\rm env}\geq\Sexc(v_i).
	\label{eq:conditional-integrated}
\end{equation}
This is a \emph{conditional two-horizon geometric bookkeeping law}, not a
consequence of the standard GSL. It depends explicitly on the additional
coarse-grained assumption in Eq.~\eqref{eq:book-second-law}.

The fixed-ADM-mass condition requires further interpretive care. It does not
identify the microscopic carrier of entropy and does not imply that the
interior excess is converted into any particular form of radiation. The
time-dependent source supporting the geometry has not yet been determined
completely, and the exterior is itself nonstationary during the collapse
\cite{Ovalle:2026wqw}. Equation~\eqref{eq:conditional-balance} should
therefore be understood as a thermodynamic consistency condition on any
future physical realization of the geometric trajectory rather than as an
independently derived flux law.

A fully dynamical entropy balance would require the true marginal or
dynamical horizon, its angular momentum, and the associated matter and
gravitational fluxes. The dynamical-horizon framework provides precisely
such balance laws in general relativity
\cite{Ashtekar:2003hk,AshtekarKrishnan:2004,Ashtekar:2025far}
and supplies the natural language for the next stage of the analysis once
the relevant axisymmetric marginal surfaces have been identified.

The stationary regular family nevertheless permits one to quantify the
largest geometric excess that such a bookkeeping law could involve. A
distinctive feature of these rotating configurations is the existence of
states for which
$h_c\simeq h$,
while $|a|\not\simeq M$.
This differs sharply from Kerr, where coincidence of the inner and outer
horizons requires \(|a|=M\). It is therefore useful to distinguish
\emph{geometric quasi-extremality}, controlled by the interior horizon
separation, from \emph{Kerr rotational extremality}, controlled by the
asymptotic charges.

For a regular quasi-extremal configuration,
\begin{align}
	\Sexc^{\rm QE}
	&\simeq
	\pi\left[(r_+^{\Kerr})^2-(r_-^{\Kerr})^2\right]
	\notag\\
	&=4\pi M\sqrt{M^2-a^2}.
	\label{eq:Sexc-QE}
\end{align}
Thus the quasi-extremal sequence approaches the supremum in
Eq.~\eqref{eq:Sexc-bound}. Relative to the Kerr two-horizon reference
value,
\begin{equation}
	\frac{\Sexc^{\rm QE}}{4\pi M^2}
	\simeq \sqrt{1-\chi^2},
	\label{eq:qe-fraction}
\end{equation}
while relative to the outer-horizon entropy,
\begin{equation}
	\frac{\Sexc^{\rm QE}}{S_+}
	\simeq
	\frac{2\sqrt{1-\chi^2}}
	{1+\sqrt{1-\chi^2}}.
	\label{eq:qe-outer-fraction}
\end{equation}
For spins that are not close to Kerr extremality these quantities can be of
order unity. Hence the regular interior can support an excess geometric
entropy of order \(M^2\) even when the asymptotic angular momentum is far
from the extremal value.

This result gives a direct geometric interpretation to the distinction
between the two notions of extremality. In the regular family the outer
surface gravity remains the ordinary finite Kerr value even as \(h_c\) is
driven close to \(h\) \cite{Ovalle:2026hair}. The large value of \(\Sexc\)
therefore originates from interior structure rather than from a
near-extremal tuning of the Kerr exterior.

There is, however, a geometric cost to approaching this entropy capacity.
The large-\(n_i\) sequence makes the radial profile increasingly sharp, and
the strict \(n_i\to\infty\) limit produces the cusp of the Kerr-mimicker
profile rather than a smooth degenerate Killing horizon. Consequently,
Eq.~\eqref{eq:Sexc-bound} should be interpreted as the upper edge of the
regular stationary state space: the supremum can be approached
arbitrarily closely by regular quasi-extremal configurations, but it is
not attained by an additional regular equilibrium black hole.

The conditional balance and the quasi-extremal capacity therefore describe
two complementary aspects of the same bookkeeping framework. The former
states how a decrease of the excess contribution would have to be
compensated in any coarse-grained thermodynamic completion, while the
latter quantifies how large that excess can become within the regular
stationary family. Together they show that the interior contribution can be
both macroscopically large and geometrically well controlled, while its
conversion into a genuine dynamical entropy flux remains a separate
physical question.

\section{Relation to previous constructions, scope, and open problems}
\label{sec:relation_scope}

The present construction is the direct rotational extension of the
inner-horizon bookkeeping proposed for regular Schwarzschild black
holes~\cite{Lobo:2026entropy}, but with one essential modification. In the
Schwarzschild endpoint there is no surviving inner horizon, so the complete
regular inner-horizon contribution is excess entropy,
\begin{equation}
	\Delta S_{\rm Schw}=\pi h_c^2.
\end{equation}
For Kerr, by contrast, the final geometry retains the reference inner
entropy \(S_-^{\Kerr}\). The subtraction in Eq.~\eqref{eq:Sexc} is therefore
precisely what is required to recover the Schwarzschild construction in the
nonrotating limit while avoiding double counting of the Kerr Cauchy horizon.

Minkowski breaking is a feature shared by the spherical and rotating
constructions; it is not the distinction between them. In the spherical
construction, loss of the Minkowski core limit and disappearance of the inner
horizon play central roles in the proposed entropy-release mechanism
\cite{Ovalle:2025dyn,Ovalle:2026schw,Casadio:2026collapse,Lobo:2026entropy}.
In the rotating $N=1$ family, Minkowski breaking occurs at $n=0$, before Kerr is reached
at $n=-1$, while the continuously tracked inner root approaches the nonzero
Kerr Cauchy-horizon radius. Thus Minkowski breaking is not equivalent to disappearance of the
inner horizon. It is the survival of the Kerr inner-horizon reference area,
not the absence of Minkowski breaking, that requires the Kerr-subtracted quantity
$\Sexc$ instead of the complete inner-horizon area. The timing $n=0$ follows
from the particular mass profile and is not asserted to be universal for all
rotating regularization mechanisms.

The de Sitter thermodynamic construction of Ref.~\cite{Lobo:2026dS} is
likewise conceptually distinct and should not be transplanted directly to
the rotating geometry. That analysis relies on a homogeneous de Sitter
region characterized by a single Hubble parameter. The present interior is
axisymmetric, contains frame dragging, and its curvature generally depends
on both \(r\) and \(\theta\). There is therefore no unique scalar Hubble
parameter whose sign could be used to define an analogous bulk entropy.
For this reason the present analysis remains deliberately tied to horizon
geometry. A genuinely rotational bulk gravitational entropy would require
an independent construction and justification.

These comparisons also clarify the scope of the present framework. The
stationary results are the strongest part of the analysis. For every regular
stationary Killing horizon, the area, angular velocity, and signed surface
gravity are well-defined geometric quantities. The Kerr-subtracted entropy,
its positivity theorem, and the bound~\eqref{eq:Sexc-bound} are exact
statements about the stationary family and do not rely on any
dynamical-horizon interpretation.

A separate level of interpretation enters when the outer and inner
horizon contributions are added. The quantity \(\Sgeom\) is used only as a
two-horizon geometric bookkeeping entropy; it is not identified with the
standard generalized entropy of a causal horizon. Accordingly, the
conditional balance introduced is intentionally
weaker than the ordinary generalized second law and should not be promoted
to a GSL statement without a microscopic or semiclassical derivation.

The time-dependent construction requires an additional distinction. The
roots \(h(v)\) and \(h_c(v)\) of \(\Delta(v,r)\) need not coincide with the
true marginal surfaces of the nonstationary axisymmetric spacetime.
Determining the relevant null expansions on nonspherical two-surfaces is
necessary before the corresponding root areas can be interpreted as genuine
dynamical-horizon entropies. The recent far-from-equilibrium extension of
black-hole thermodynamics based on dynamical horizon segments makes this
identification particularly timely~\cite{Ashtekar:2025far}. Until such an
analysis is carried out, the time-dependent quantities introduced here
should be understood as exact horizon-root diagnostics rather than as a
derived dynamical entropy law.

Closely related to this issue is the physical source of the evolving
geometry. Before \(\Sexcdot\) can be associated with matter entropy,
gravitational-wave entropy, or any other flux, the stress-energy content of
the time-dependent spacetime must be identified explicitly. The exact
outer-root identity and inner-root transport law determine the geometric
evolution of the roots, but they do not by themselves specify the
thermodynamic reservoir into which the excess contribution is transferred.

The Kerr-forming interpolation also illustrates a further limitation. The
continuous extension of the mass profile through \(n=2\) removes the
apparent algebraic pole but does not remove the curvature singularity.
Indeed, the trajectory necessarily leaves the regular-black-hole sector
before reaching the Kerr endpoint. {Local integrability of the Ricci-scalar
density does not by itself establish physical viability or traversability}~\cite{Lukash:2013ts,
	Ovalle:2023dtn,Arrechea:2025vcc,Casadio:2026collapse}. In particular, the
finiteness of \(\Sexc\) across such a regime should not be interpreted as a
regularity criterion. The statement \(\Sexc\to0\) refers to the Kerr endpoint
of the interpolating trajectory, not to a continuous path entirely contained
within the space of regular black holes.
An exact Schwarzschild or Kerr endpoint does not retain the regular
Minkowski core limit. Nevertheless, this endpoint fact alone does not prove
that every possible interpolation must undergo Minkowski breaking at a finite parameter value.
Alternative profiles might preserve that limit at all finite stages and lose
it only in a nonuniform singular endpoint limit. Whether a physically
consistent regularization can postpone Minkowski breaking in this way is an open question;
its existence or universality is not established here.

An independent issue is the stability of the inner horizon. The existence of
a stationary Cauchy horizon does not imply nonlinear stability. Classical
mass-inflation analyses show that small counterstreaming perturbations can
turn an otherwise regular Cauchy horizon into a singular boundary
\cite{Poisson:1989zz,Ori:1991zz,Ori:1992,Sbierski:2023}. For regular black
holes, both generic inner-horizon instability and mechanisms proposed to
soften or evade the standard exponential mass-inflation behavior have been
investigated~\cite{CarballoRubio:2021,CarballoRubio:2022,Bonanno:2025ch,
	Frolov:2026qg}. The outcome is therefore model dependent. The
\(C^0\)-stability result for perturbed Kerr further emphasizes that
continuous extendibility and curvature or derivative instability are
logically distinct notions~\cite{DafermosLuk:2025}. None of these results
establishes stability of the rotating family considered here. Any physical
interpretation of the stationary discrete family of geometric entropy
	values must therefore be supplemented by a dedicated perturbative analysis
of the corresponding interior states.

Finally, the microscopic interpretation of \(\Sexc\) remains open. Its most
robust meaning at present is geometric: it measures the excess of the
regular inner-horizon area over the Kerr inner-horizon area at the same
asymptotic charges \((M,J)\). Because the stationary regular family is
labelled by discrete interior integers, this quantity takes a discrete set
of geometric values even when the exterior charges are held fixed. This
structure is suggestive of an underlying state-counting interpretation, but
no microscopic entropy spectrum follows from the classical area law alone.
Establishing such an interpretation would require an independent quantum
description of the interior degrees of freedom.

Taken together, these observations delineate the precise status of the
construction. The Kerr-subtracted excess entropy is an exact and controlled
stationary geometric diagnostic, with a well-defined Schwarzschild limit
and a natural Kerr reference state. Its extension along the prescribed
time-dependent trajectory provides useful root-based bookkeeping, but a
fully physical thermodynamic interpretation ultimately requires three
further ingredients: identification of the true axisymmetric marginal
surfaces, determination of the dynamical stress-energy and gravitational
fluxes, and an analysis of the stability of the inner-horizon sector.

\section{Discussion and conclusions}
\label{sec:conclusions}

We have developed a thermodynamic framework for a class of regular rotating
black holes and for the exact geometric model describing their final
approach to Kerr. The central conceptual point is that the rotational
problem has a nonzero inner-horizon baseline. A Kerr black hole retains a
Cauchy horizon, so the complete area of the regular inner horizon is not an
entropy contribution that disappears at the endpoint. The appropriate
charge-matched measure is instead the excess
\(\Sexc=\pi[h_c^2-(r_-^{\Kerr})^2]\).

This subtraction leads to a remarkably clean stationary structure. For the
regular mass profiles satisfying the hypotheses of Theorem~\ref{thm:positive},
the regular Cauchy horizon lies strictly
outside the Kerr inner horizon with the same \(M\) and \(J\), implying
\(\Sexc>0\). The excess is bounded above by
\(4\pi M\sqrt{M^2-a^2}\), reduces exactly to the hidden Schwarzschild
inner-horizon entropy when rotation is removed, and tends to zero at the
Kerr endpoint. The two-horizon Kerr entropy sum is independent of spin at
fixed mass, so ordinary Kerr rotation only redistributes entropy between the
outer and inner reference horizons; \(\Sexc\) isolates the additional
interior contribution.

The same quantity controls the deformation of the Kerr entropy product.
This provides a useful link to the broader literature on inner-horizon
mechanics and area products, while also clarifying why no contradiction
arises with electrovacuum universality results: the present interior is
supported by an anisotropic nonvacuum source. More physically, the excess
entropy distinguishes different interior geometries that possess identical
primary asymptotic charges and the same Kerr exterior. It therefore gives a
thermodynamic diagnostic of hidden interior state information rather than a
new conserved hair.

The quasi-extremal sector makes this interpretation especially sharp.
Regular configurations can have \(h_c\) arbitrarily close to the outer Kerr
radius even for spin far from \(|a|=M\), allowing an order-\(M^2\) excess
entropy without Kerr rotational extremality. The corresponding upper bound
is only a supremum: exact saturation at fixed nonextremal spin occurs in the
mimicker limit, where the nominal horizon develops a cusp and ceases to be
a regular Killing horizon. Large hidden entropy is therefore associated
with approaching the edge of the regular interior state space.

For the time-dependent construction we derived the exact inner-root
transport equation and the corresponding evolution law for \(\Sexc\). A
particularly important technical point is that the apparent \(1/(n-2)\)
pole of the minimal mass profile is removable: at \(n=2\) the mass function
has the finite logarithmic form \(M x^3(1-3\ln x)\). The horizon-root
equation and entropy diagnostic can therefore be continued through this
value. {The corresponding stationary-profile curvature, however, is singular there}, so the continuous
root evolution must not be confused with a continuous family of regular
black holes. The Kerr-forming trajectory necessarily passes through a
singular intermediate regime before reaching \(n=-1\).
The stationary profiles have a {locally absolutely integrable Ricci-scalar density
for $-1<n\leq2$; the Ricci-flat Kerr endpoint at $n=-1$ is treated separately}. Minkowski breaking occurs at $n=0$,
before that endpoint and independently of inner-root disappearance.
These corrections leave the stationary excess-entropy subtraction,
positivity theorem, bound, and Kerr redistribution identities unchanged.

We also separated the proposed entropy bookkeeping from the standard GSL.
If a physical completion admits a nondecreasing coarse-grained quantity
\(S_{\rm env}+S_++S_c\), then the decrease of \(\Sexc\) imposes a lower bound
on the compensating environmental entropy. This is a conditional statement,
not an independent theorem of semiclassical gravity. Establishing a true
dynamical entropy law requires the actual axisymmetric marginal surface,
the associated angular momentum, and the matter and gravitational fluxes.

Several concrete next steps follow. The first is to solve the dimensionless
root equation along representative collapse trajectories and determine
whether \(\Sexc(v)\) is monotonic or exhibits intermediate storage phases.
The second is to locate the true dynamical horizon and compare its area
balance with the horizon-root diagnostic developed here. The third is to
identify the time-dependent stress tensor and connect changes of the excess
entropy to physical fluxes. Finally, the stability of the inner horizon must
be assessed to determine which stationary entropy levels can survive
perturbations for physically meaningful times.

The broader conclusion is that rotating regular black holes possess a
well-defined internal geometric sector that is invisible to the primary
charges \((M,J)\). Kerr supplies the natural reference state, and its
surviving inner horizon provides the subtraction required to isolate this
sector. The resulting excess entropy is positive, bounded, analytically
controlled, and compatible with both the Schwarzschild limit and the Kerr
endpoint. It therefore offers a precise bridge between regular black-hole
interiors, inner-horizon mechanics, rotational collapse, and the
thermodynamic bookkeeping of hidden gravitational structure.

\acknowledgments{
	MER thanks Conselho Nacional de Desenvolvimento Cient\'ifico e Tecnol\'ogico - CNPq, Brazil, for partial financial support. FSNL acknowledges support from the Funda\c{c}\~{a}o para a Ci\^encia e a Tecnologia (FCT) Scientific Employment Stimulus contract with reference CEECINST/00032/2018, and funding through the research grants UIDB/04434/2020, UIDP/04434/2020 and PTDC/FIS-AST/0054/2021. JO is partially supported by ANID-FONDECYT Grant No. 1250227.}
	
\appendix
\section{Local integrability of the Ricci-scalar density and Minkowski breaking}
\label{app:integrability-MB}

For the stationary profiles of the minimal model~\eqref{eq:m-mu}, with
$M$ and $a$ fixed, the Ricci scalar for $n\neq2$ is
\begin{equation}
 R(r,\theta)=\frac{6M}{\rho^2h}\frac{n+1}{n-2}
 \left[4\left(\frac rh\right)^2-(n+2)\left(\frac rh\right)^n\right].
 \label{app:1}
\end{equation}
Since $\sqrt{-g}=\rho^2\sin\theta$, the potentially singular factor $\rho^{-2}$
cancels in the invariant integration density. For a finite stationary-time
interval of duration $\Delta t$ and $0<\epsilon\ll h$, define
\begin{equation}
 I(\epsilon)=\int_{t_0}^{t_0+\Delta t}\!dt
 \int_0^\epsilon\!dr\int_0^\pi\!d\theta\int_0^{2\pi}\!d\phi\,
 R\sqrt{-g}.
 \label{app:integral}
\end{equation}
For $-1<n<2$, direct integration gives
\begin{align}
 I(\epsilon)
 &=\frac{24\pi M\Delta t}{n-2}
 \left[\frac{4(n+1)}{3}\left(\frac\epsilon h\right)^3
 -(n+2)\left(\frac\epsilon h\right)^{n+1}\right]
 \notag\\
 &\sim-\frac{24\pi M\Delta t(n+2)}{n-2}
 \left(\frac\epsilon h\right)^{n+1}.
 \label{app:integral-evaluated}
\end{align}
Both powers are integrable for $n>-1$ in this interval. The same power-counting
applies to $|R|\sqrt{-g}$, so the result establishes local absolute
integrability of the {Ricci-scalar density}, not merely cancellation in a signed
integral. This conclusion is for each fixed $n$; it does not imply uniform
convergence as $n\to-1$.

The endpoints must be evaluated separately. At $n=-1$, the mass profile is
$m=M$ and $R=0$ for $r>0$: this is the Ricci-flat Kerr endpoint, whose ring
singularity is not removed. At $n=2$, the continued profile in
Eq.~\eqref{eq:mu-limit} gives
\[
 R\big|_{n=2}=-\frac{18M}{\rho^2h}
 \left(\frac rh\right)^2\left[1+4\ln\left(\frac rh\right)\right],
\]
and
\[
 I(\epsilon)\big|_{n=2}=8\pi M\Delta t
 \left(\frac\epsilon h\right)^3
 \left[1-12\ln\left(\frac\epsilon h\right)\right].
\]
The logarithmic {Ricci-scalar density} is also locally absolutely integrable. {Thus,
within the singular stationary-profile sector traversed by the trajectory,
local absolute integrability of the Ricci-scalar density holds for $-1<n\leq2$.
The endpoint $n=-1$ is separately Ricci-flat for $r>0$.} This {test of Ricci-scalar density integrability}
does not establish integrability of {other curvature invariants, including the Kretschmann scalar}, a
distributional extension through the Kerr ring, or regularity of the
nonstationary spacetime.

For Minkowski breaking, the stationary metric~\eqref{eq:stationary-metric} gives
$g_{tt}(r,\pi/2)=-1+2m(r)/r$. The core limit is
\begin{equation}
 \lim_{r\to0^+}g_{tt}(r,\pi/2)=
 \begin{cases}
 -1,&n>0,\\[2pt]
 -1+3M/h,&n=0,\\[2pt]
 +\infty,&-1\leq n<0.
 \end{cases}
 \label{MBx}
\end{equation}
The $n>0$ case includes the logarithmic continuation at $n=2$.
With the asymptotically normalized stationary time coordinate, $g_{tt}$ is
the norm of the stationary Killing vector. Equation~\eqref{MBx} therefore
identifies the loss of its Minkowski core limit at $n=0$, already within the
singular interval. In particular, at $n=0$ the equatorial limit differs from
the value $-1$ obtained at fixed nonequatorial angles. The condition
$g_{tt}\to-1$ alone is not a criterion for bounded curvature: the profiles
with $0<n\leq2$ illustrate this distinction.

Minkowski breaking is therefore diagnosed without invoking disappearance of a Cauchy
horizon. Applied to the instantaneous stationary profiles of the prescribed
interpolation, this criterion is crossed at the finite parameter value
$n=0$, before the asymptotic Kerr endpoint $n=-1$. Its timing and its
relation to physical entropy flux require separate dynamical analysis.


\begin{thebibliography}{99}

\bibitem{Bekenstein:1973ur}
J.~D.~Bekenstein,
``Black holes and entropy,''
Phys. Rev. D \textbf{7} (1973), 2333--2346.

\bibitem{Hawking:1975vcx}
S.~W.~Hawking,
``Particle creation by black holes,''
Commun. Math. Phys. \textbf{43} (1975), 199--220;
Erratum: Commun. Math. Phys. \textbf{46} (1976), 206.

\bibitem{Bardeen:1973gs}
J.~M.~Bardeen, B.~Carter and S.~W.~Hawking,
``The four laws of black hole mechanics,''
Commun. Math. Phys. \textbf{31} (1973), 161--170.

\bibitem{Wald:1993nt}
R.~M.~Wald,
``Black hole entropy is the Noether charge,''
Phys. Rev. D \textbf{48} (1993), R3427--R3431,
[arXiv:gr-qc/9307038].

\bibitem{Iyer:1994ys}
V.~Iyer and R.~M.~Wald,
``Some properties of Noether charge and a proposal for dynamical black hole entropy,''
Phys. Rev. D \textbf{50} (1994), 846--864,
[arXiv:gr-qc/9403028].

\bibitem{Jacobson:1993xs}
T.~Jacobson, G.~Kang and R.~C.~Myers,
``On black hole entropy,''
Phys. Rev. D \textbf{49} (1994), 6587--6598,
[arXiv:gr-qc/9312023].

\bibitem{Kerr:1963ud}
R.~P.~Kerr,
``Gravitational field of a spinning mass as an example of algebraically special metrics,''
Phys. Rev. Lett. \textbf{11} (1963), 237--238.

\bibitem{Carter:1968rr}
B.~Carter,
``Global structure of the Kerr family of gravitational fields,''
Phys. Rev. \textbf{174} (1968), 1559--1571.

\bibitem{Teukolsky:2014vca}
S.~A.~Teukolsky,
``The Kerr metric,''
Class. Quant. Grav. \textbf{32} (2015), 124006,
[arXiv:1410.2130 [gr-qc]].

\bibitem{Curir:1979}
A.~Curir,
``Spin entropy of a rotating black hole,''
Nuovo Cim. B \textbf{51} (1979), 262--266.

\bibitem{CurirFrancaviglia:1979}
A.~Curir and M.~Francaviglia,
``Spin thermodynamics of a Kerr black hole,''
Nuovo Cim. B \textbf{52} (1979), 165--176.

\bibitem{OkamotoKaburaki:1992}
I.~Okamoto and O.~Kaburaki,
``The `inner-horizon thermodynamics' of Kerr black holes,''
Mon. Not. Roy. Astron. Soc. \textbf{255} (1992), 539--544.

\bibitem{Castro:2012av}
A.~Castro and M.~J.~Rodriguez,
``Universal properties and the first law of black hole inner mechanics,''
Phys. Rev. D \textbf{86} (2012), 024008,
[arXiv:1204.1284 [hep-th]].

\bibitem{Castro:2013pqa}
A.~Castro, N.~Dehmami, G.~Giribet and D.~Kastor,
``On the universality of inner black hole mechanics and higher curvature gravity,''
JHEP \textbf{07} (2013), 164,
[arXiv:1304.1696 [hep-th]].

\bibitem{Cvetic:2010mn}
M.~Cveti\v{c}, G.~W.~Gibbons and C.~N.~Pope,
``Universal area product formulae for rotating and charged black holes in four and higher dimensions,''
Phys. Rev. Lett. \textbf{106} (2011), 121301,
[arXiv:1011.0008 [hep-th]].

\bibitem{Chen:2012mh}
B.~Chen, S.~x.~Liu and J.~j.~Zhang,
``Thermodynamics of black hole horizons and Kerr/CFT correspondence,''
JHEP \textbf{11} (2012), 017,
[arXiv:1206.2015 [hep-th]].

\bibitem{Poisson:1989zz}
E.~Poisson and W.~Israel,
``Inner-horizon instability and mass inflation in black holes,''
Phys. Rev. Lett. \textbf{63} (1989), 1663--1666.

\bibitem{Ori:1991zz}
A.~Ori,
``Inner structure of a charged black hole: An exact mass-inflation solution,''
Phys. Rev. Lett. \textbf{67} (1991), 789--792.

\bibitem{Ori:1992}
A.~Ori,
``Structure of the singularity inside a realistic rotating black hole,''
Phys. Rev. Lett. \textbf{68} (1992), 2117--2120.

\bibitem{Sbierski:2023}
J.~Sbierski,
``Instability of the Kerr Cauchy Horizon Under Linearised Gravitational Perturbations,''
Annals PDE \textbf{9} (2023), 7.

\bibitem{DafermosLuk:2025}
M.~Dafermos and J.~Luk,
``The interior of dynamical vacuum black holes I: The $C^0$-stability of the Kerr Cauchy horizon,''
Annals Math. \textbf{202} (2025), 309--630,
[arXiv:1710.01722 [gr-qc]].

\bibitem{Dymnikova:1992}
I.~Dymnikova,
``Vacuum nonsingular black hole,''
Gen. Rel. Grav. \textbf{24} (1992), 235--242.

\bibitem{Hayward:2006}
S.~A.~Hayward,
``Formation and evaporation of nonsingular black holes,''
Phys. Rev. Lett. \textbf{96} (2006), 031103,
[arXiv:gr-qc/0506126].

\bibitem{Bambi:2013ufa}
C.~Bambi and L.~Modesto,
``Rotating regular black holes,''
Phys. Lett. B \textbf{721} (2013), 329--334,
[arXiv:1302.6075 [gr-qc]].

\bibitem{Casadio:2023iqt}
R.~Casadio, A.~Giusti and J.~Ovalle,
``Quantum rotating black holes,''
JHEP \textbf{05} (2023), 118,
[arXiv:2303.02713 [gr-qc]].

\bibitem{Lukash:2013ts}
V.~N.~Lukash and V.~N.~Strokov,
``Space-times with integrable singularity,''
Int. J. Mod. Phys. A \textbf{28} (2013), 1350007,
[arXiv:1301.5544 [gr-qc]].

\bibitem{Ovalle:2023dtn}
J.~Ovalle,
``Black holes without Cauchy horizons and integrable singularities,''
Phys. Rev. D \textbf{107} (2023), 104005,
[arXiv:2305.00030 [gr-qc]].

\bibitem{Arrechea:2025vcc}
J.~Arrechea, S.~Liberati, H.~Neshat and V.~Vellucci,
``Physical and theoretical challenges to integrable singularities,''
Phys. Rev. D \textbf{112} (2025), 044024,
[arXiv:2504.17863 [gr-qc]].

\bibitem{Ovalle:2024vuk}
J.~Ovalle,
``Schwarzschild black hole revisited: Before the complete collapse,''
Phys. Rev. D \textbf{109} (2024), 104032,
[arXiv:2405.06731 [gr-qc]].

\bibitem{Ovalle:2025dyn}
J.~Ovalle,
``Interior evolution of regular Schwarzschild black holes,''
    Phys. Lett. B \textbf{881} (2026), 140912,
[arXiv:2509.00816 [gr-qc]].

\bibitem{Ovalle:2026schw}
J.~Ovalle, R.~Casadio and A.~Kamenshchik,
``Schwarzschild black hole singularity formation,''
Phys. Rev. D \textbf{113} (2026), 064042,
[arXiv:2603.06451 [gr-qc]].

\bibitem{Ovalle:2026hair}
J.~Ovalle,
``Kerr black holes without primary hairs,''
[arXiv:2607.14910 [gr-qc]].


\bibitem{Lobo:2026entropy}
F.~S.~N.~Lobo and M.~E.~Rodrigues,
``Entropy release from Minkowski breaking in regular Schwarzschild black holes,''
[arXiv:2607.14079 [gr-qc]].

\bibitem{Lobo:2026dS}
F.~S.~N.~Lobo and M.~E.~Rodrigues,
``Entropy dynamics in gravitational collapse: From Minkowski breaking to de Sitter thermodynamics,''
[arXiv:2607.24349 [gr-qc]].

\bibitem{Ovalle:2026wqw}
J.~Ovalle,
``Formation of the Kerr black hole: an exact model,''
[arXiv:2608.06519 [gr-qc]].

\bibitem{Hayward:1993wb}
S.~A.~Hayward,
``General laws of black-hole dynamics,''
Phys. Rev. D \textbf{49} (1994), 6467--6474,
[arXiv:gr-qc/9303006].

\bibitem{Ashtekar:2001is}
A.~Ashtekar, C.~Beetle and J.~Lewandowski,
``Mechanics of rotating isolated horizons,''
Phys. Rev. D \textbf{64} (2001), 044016,
[arXiv:gr-qc/0103026].

\bibitem{Ashtekar:2003hk}
A.~Ashtekar and B.~Krishnan,
``Dynamical horizons and their properties,''
Phys. Rev. D \textbf{68} (2003), 104030,
[arXiv:gr-qc/0308033].

\bibitem{AshtekarKrishnan:2004}
A.~Ashtekar and B.~Krishnan,
``Isolated and dynamical horizons and their applications,''
Living Rev. Relativity \textbf{7} (2004), 10,
[arXiv:gr-qc/0407042].

\bibitem{Booth:2005qc}
I.~Booth,
``Black hole boundaries,''
Can. J. Phys. \textbf{83} (2005), 1073--1099,
[arXiv:gr-qc/0508107].

\bibitem{Ashtekar:2025far}
A.~Ashtekar, D.~E.~Paraizo and J.~Shu,
``Thermodynamics of black holes, far from equilibrium,''
Phys. Rev. Lett. \textbf{136} (2026), 251405,
[arXiv:2512.11659 [gr-qc]].

\bibitem{Gao:2001ut}
S.~Gao and R.~M.~Wald,
``The `physical process' version of the first law and the generalized second law for charged and rotating black holes,''
Phys. Rev. D \textbf{64} (2001), 084020,
[arXiv:gr-qc/0106071].

\bibitem{Wall:2011hj}
A.~C.~Wall,
``A proof of the generalized second law for rapidly changing fields and arbitrary horizon slices,''
Phys. Rev. D \textbf{85} (2012), 104049;
Erratum: Phys. Rev. D \textbf{87} (2013), 069904,
[arXiv:1105.3445 [gr-qc]].

\bibitem{Gurses:1975vu}
M.~G\"urses and F.~G\"ursey,
``Lorentz covariant treatment of the Kerr--Schild metric,''
J. Math. Phys. \textbf{16} (1975), 2385--2390.

\bibitem{Ansorg:2009yi}
M.~Ansorg and J.~Hennig,
``Inner Cauchy horizon of axisymmetric and stationary black holes with surrounding matter in Einstein--Maxwell theory,''
Phys. Rev. Lett. \textbf{102} (2009), 221102,
[arXiv:0903.5405 [gr-qc]].

\bibitem{Visser:2012wu}
M.~Visser,
``Area products for stationary black hole horizons,''
Phys. Rev. D \textbf{88} (2013), 044014,
[arXiv:1205.6814 [hep-th]].

\bibitem{Pradhan:2016}
P.~Pradhan,
``Area (or entropy) product formula for a regular black hole,''
Gen. Rel. Grav. \textbf{48} (2016), 19,
[arXiv:1512.06187 [gr-qc]].

\bibitem{Casadio:2026collapse}
R.~Casadio, A.~Giusti, A.~Kamenshchik and J.~Ovalle,
``On gravitational collapse and integrable singularities,''
[arXiv:2605.01808 [gr-qc]].

\bibitem{CarballoRubio:2021}
R.~Carballo-Rubio, F.~Di~Filippo, S.~Liberati, C.~Pacilio and M.~Visser,
``Inner horizon instability and the unstable cores of regular black holes,''
JHEP \textbf{05} (2021), 132,
[arXiv:2101.05006 [gr-qc]].

\bibitem{CarballoRubio:2022}
R.~Carballo-Rubio, F.~Di~Filippo, S.~Liberati, C.~Pacilio and M.~Visser,
``Regular black holes without mass inflation instability,''
JHEP \textbf{09} (2022), 118,
[arXiv:2205.13556 [gr-qc]].


\bibitem{Bonanno:2025ch}
A.~Bonanno, A.~Panassiti and F.~Saueressig,
``Cauchy horizon stability and instability of regular black holes,''
    Phys. Rev. D \textbf{114} (2026), 064027,
[arXiv:2507.03581 [gr-qc]].

\bibitem{Frolov:2026qg}
V.~P.~Frolov and A.~Zelnikov,
``Regular black holes in quasitopological gravity: Null shells and mass inflation,''
Phys. Rev. D \textbf{113} (2026), 084007,
[arXiv:2601.01861 [gr-qc]].


\end{thebibliography}
\end{document}